\documentclass[10pt,dvips]{article}
\usepackage{amssymb,amsfonts,amsmath,latexsym}
\usepackage{graphics,graphicx,epsf}
\newcommand{\be}{\begin{equation}}
\newcommand{\ee}{\end{equation}}
\newcommand{\bea}{\begin{eqnarray}}
\newcommand{\eea}{\end{eqnarray}}
\newcommand{\ba}{\begin{array}}
\newcommand{\ea}{\end{array}}
\newcommand{\bt}{\begin{tabular}}
\newcommand{\et}{\end{tabular}}

\newcommand{\fr}{\frac}
\newcommand{\ci}{\cite}
\newcommand{\cl}{\centerline}
\newcommand{\bs}{\bigskip}

\newcommand{\vs}{\vspace}

\newcommand{\en}{\eqno}

\newcommand{\fns}{\footnotesize}

\newcommand{\bbib}{\begin{thebibliography}}
\newcommand{\ebib}{\end{thebibliography}}

\newcommand{\xra}{\xrightarrow}
\newcommand{\und}{\underline}

\begin{document}
\bs
\cl{\bf LARGE LINEAR MAGNETORESISTIVITY }
\cl{\bf IN STRONGLY INHOMOGENEOUS}
\cl{\bf PLANAR AND LAYERED SYSTEMS}

\bs

\cl{\bf S.A.Bulgadaev \footnote{e-mail: bulgad@itp.ac.ru},
\bf F.V.Kusmartsev \footnote{e-mail: F.Kusmartsev@lboro.ac.uk}}

\bs \cl{\fns Landau Institute for Theoretical Physics,
Chernogolovka, Moscow Region, Russia, 142432} \cl{\fns
Department of Physics, Loughborough University, Loughborough, LE11
3TU, UK}

\bs

\begin{quote}
\footnotesize{ Explicit expressions for magnetoresistance $R$ of
planar and layered strongly inhomogeneous two-phase systems are
obtained, using exact dual transformation, connecting effective
conductivities of in-plane isotropic two-phase systems with and
without magnetic field. These expressions allow to describe the
magnetoresistance of various inhomogeneous media at arbitrary
concentrations $x$ and magnetic fields $H$. All expressions show
large linear magnetoresistance effect with different dependencies
on the phase concentrations. The corresponding plots of the $x$- and
$H$-dependencies of $R(x,H)$ are represented for various
values, respectively, of magnetic field and concentrations at some
values of inhomogeneity parameter. The obtained results show a
remarkable similarity with the existing experimental data on
linear magnetoresistance in silver chalcogenides
$Ag_{2+\delta}Se.$ A possible physical explanation of this
similarity is proposed. It is shown that the random, stripe type, structures of inhomogeneities are the most suitable for a fabrication
of magnetic sensors and a storage of information at room temperatures.}
\end{quote}

\bs
\cl{PACS: 75.70.Ak, 72.80.Ng, 72.80.Tm, 73.61.-r}
\bs

\underline{\bf 1. Introduction}

\bs

It was established recently that new materials, such as oxides and
chalcogenides,  have often unusual magneto-transport properties.
For example, the magnetoresistance becomes very large (the so
called colossal magnetoresistance) in manganites \ci{1} or grows
approximately linearly with magnetic field up to very high fields
in silver chalcogenides \ci{2}. The large linear magnetoresistance (LLMR)
takes place in
thin films of $Ag_{2+\delta}Se,Te$ in a wide region of temperatures, from 
low ($\sim 1 K$) till room temperatures ($\sim 300 K$). 
At the moment there exist two approaches in a theoretical explanation of a linear behaviour of the magnetoresistance. The first, a quantum one, is proposed by 
Abrikosov and is based on the quantum theory of possible changes of spectrum properties of  semimetals or narrow gap semiconductors \ci{3}. It can be applied, for example, for an explanation of low temperature properties of $Ag_{2+\delta}Te$ (\ci{4}). The second approach is pure classical and is based on
the importance of the phase inhomogeneities, which take place in
these materials on small (till nanometer) scales, for an existence of the LLMR effect at moderate temperatures \ci{2}. This approach is applicable at moderate temperatures and for systems where a mean size or some characteristic size of inhomogeneities $l_c \gg l_0$ (here $l_0$ is a free path length of the charge carriers). In the framework of the second approach Parish and Littlewood have proposed the network model constructed from conducting discs and have shown by its numerical simulation that the LLMR appears in this model, when the parameters of a
disc's impedance (in particular, a mobility $\mu$) are random and have a continuous
wide distribution with $\langle \mu \rangle = 0$ \ci{5,6}. This model is similar to the usual wire network model \ci{7}, but, in order to describe a
dependence of the effective resistivity on magnetic field, it uses as  building
blocks the four terminal discs instead of wires. 

In the framework of the classical  approach there is another
possibility to describe magneto-transport properties of inhomogeneous planar (or
layered, inhomogeneous in  planes, but constant in the direction orthogonal to planes) systems in perpendicular magnetic field. It is connected with an existence of the exact dual transformation, relating the effective conductivity $\hat \sigma_e$
(and the effective resistivity $\hat \rho_e = \hat \sigma_e^{-1}$)
of planar inhomogeneous self-dual two-phase systems without and
with magnetic field \ci{8,9}. The existence of this transformation
is a direct consequence of the exact Keller - Dykhne duality of
two-dimensional systems \ci{10,11}. 

In this letter, using this exact transformation and the known expressions for $\hat \sigma_e$ of three inhomogeneous models with different random
 structures from \ci{12}, we will give the explicit approximate expressions
for the effective resistivity $\hat \rho_e$ of self-dual two-phase systems applicable at arbitrary values of phase concentrations and magnetic fields and in a wide region of partial conductivities.
We will present also the $x$- and $H$-dependencies plots of the magnetoresistance $R(x,H)$ at some characteristic values, respectively, of magnetic field $H$ or phase concentrations. These plots unambiguously show the existence of the large linear
magnetoresistance effect in these classical systems. A comparison of results,  obtained here analytically, with the known experimental data on the magnetoresistance behaviour in silver chalcogenides $Ag_{2+\delta}Se$ demonstrates their remarkable qualitative similarity. A physical explanation of this
similarity is proposed. A possibility of an application of random, stripe type, inhomogeneities for a construction of magnetic sensors and magnetic read-write technologies is indicated.

\bs
\und{\bf 2. Effective resistivity  in magnetic field}
\bs

The effective conductivity of two-phase isotropic systems in
magnetic field has the following form
$$
\hat \sigma = \sigma_{ik} = \sigma_d \delta_{ik} + \sigma_t
\epsilon_{ik}, \quad \sigma_d ({\bf H}) = \sigma_d (-{\bf H}),
\quad \sigma_t ({\bf H}) = -\sigma_t (-{\bf H}),
\en(1)
$$
here $\delta_{ik}$ is the Kronecker symbol, $\epsilon_{ik}$ is the unit antisymmetric tensor.
The effective resistivity $\hat \rho_{e}$ is defined by the inverse matrix
$$
\hat \rho = \rho_{ik} = \rho_d \delta_{ik} + \rho_t
\epsilon_{ik}, \quad \rho_d ({\bf H}) = \rho_d (-{\bf H}),
\quad \rho_t ({\bf H}) = -\rho_t (-{\bf H}),
\en(2)
$$
where
$$
\rho_d = \sigma_d/(\sigma_d^2 + \sigma_t^2), \quad \rho_t =
-\sigma_t/(\sigma_d^2 + \sigma_t^2). \en(3)
$$

The effective resistivity $\hat \rho_e$ and, consequently,
$\rho_{ed}, \; \rho_{et}$ (we assume here that
$\rho_{id} \ge 0$) for self-dual systems must be a symmetric
function of pairs of partial arguments ($\hat \rho_i, x_i$) and a homogeneous
(a degree 1) function of $\rho_{di,ti}.$ For this reason it is
invariant under permutation of pairs of partial parameters
$$
\hat \rho_{e}(\hat \rho_1, x_1|\hat \rho_2, x_2) = \hat
\rho_{e}(\hat \rho_2, x_2|\hat \rho_1, x_1).
\en(4)
$$
The effective resistivity of inhomogeneous systems must also
reduce to some partial $\hat \rho_i$, when $x_i = 1 \;(i=1,2).$

In our previous paper \ci{13} we have obtained explicit expressions for
effective conductivities of planar inhomogeneous self-dual systems in magnetic field and
have shown that they have properties qualitatively compatible with
the large linear magnetoresistance effect. Since the people
usually mesure in experiments the effective resistivities and such
value as the magnetoresistance, we consider in this letter  the properties of the
$\rho_{ed}(x,H)$ and the
magnetoresistance $R(x,H)$ of strongly inhomogeneous two-phase planar (and
layered) systems. The investigation of properties of the classical Hall effect in these systems will be done in the subsequent paper.
The magnetoresistance of two-dimensional inhomogeneous media  is determined as follows
$$
R(x,H) = (\rho_{ed}(x,H) - \rho_{ed}(x,0))/\rho_{ed}(x,0).
\en(5)
$$
One needs to note here that there is another definition of the magnetic resistance
in the literature, which differs from (5) by the normalization
$$
R_1(x,H) = (\rho_{ed}(x,H) - \rho_{ed}(x,0))/\rho_{ed}(x,H).
\en(5')
$$
We will use here a definition (5), since it allows to show in a more obvious way
the LLMR effect.  
For isotropic systems without magnetic field $\rho_{ed}(x,0) \equiv \rho_{e0}= (\sigma_{e0})^{-1},$ where $\sigma_{e0}$ is the effective conductivity at $H=0,$ and, for self-dual systems, $\rho_{e0}$ has the same functional form as $\sigma_{e0}.$ 
Then one can write
$R$ as
$$
R(x,H) = \sigma_{e0}\rho_{ed}(x,H) - 1 = \sigma_{e0} \fr{\sigma_{ed}} {\sigma_{ed}^2 + \sigma_{et}^2} - 1.
\en(6)
$$
It follows from (5) that $R(x,0)= 0$ and can be nonzero
only for $H\ne 0.$ Thus, it describes only transport properties connected with
magnetic field. Note that for large values $R$ can be approximated by
$\sigma_{e0}\rho_{ed}.$ 
In order to describe the dependence of the $R(x,H)$ of
inhomogeneous system on magnetic field one must know also the
functional dependences on $H$ of partial conductivities
$\sigma_{id}({\bf H}), \sigma_{it}({\bf H}), \; (i=1,2).$ They
usually can be approximated by the standart (metallic type)
formulae \ci{11}
$$
\sigma_{id}({\bf H}) = \fr{\sigma_{i0}}{1+ \beta_i^2}, \quad
\sigma_{it}({\bf H}) = \fr{\sigma_{i0} \beta_i}{1+ \beta_i^2}, \quad
\beta_i = \mu_{i} H, \quad i=1,2,
\en(7)
$$
where $\sigma_{i0}$ are the partial conductivities of phases
without magnetic field, $\mu_i = e_i \tau_i/m_i$ are the corresponding mobilities,
$\tau_i, e_i, m_i$ are, respectively,  carrier's times of life, charges and masses. Thus $R(x,H)$ depends also on 4 partial parameters: $\sigma_i, \;\mu_i.$
Since $\sigma_{ed,t}$ are homogeneous functions of $\sigma_{i0}$
and the mobilities enter always in a combination with magnetic field,
really $R(x,H)$ depends (besides $x$ and renormalized magnetic
field $H' = \mu_1 H$, below we will assume that $\mu_1 = 1$), on two inhomogeneity parameters: $z,$
connected with an inhomogeneity of conductivities, and $\eta,$
connected with an inhomogeneity of mobilities. They may be chosen as
the corresponding ratios (we choose also that $\sigma_{20}/\sigma_{10} \le 1$)
$$
z = \sigma_{20}/\sigma_{10} \quad (0 \le z \le 1), \quad \eta = \mu_2/\mu_1 \quad (-\infty \le \eta \le \infty). 
\en(8)
$$

It is useful also to note that $\rho_{d}$ for homogeneous systems with
conductivities from (7) has the form
$
\rho_{id}(H) = \sigma_{0i}^{-1}, \quad i=1,2,
$
and, consequently, for these homogeneous systems
$$
R(x,H) = \sigma_{ed}(x,0) \rho_{ed}(x,H) - 1  \xra[x=0,1]{} 0.
\en(9)
$$
The equation (9) defines the boundary values of $R(x,H)$ at $x=0,1$  for inhomogeneous systems, satisfying the representation (7).

\bs

\und{\bf 3. Magnetoresistance of planar inhomogeneous systems.}

\bs

Recently, using exact "magnetic" duality transformation,
connecting the effective conductivities of inhomogeneous systems
with and without magnetic field \ci{9}, we have obtained  three
explicit approximate expressions for the effective conductivity
$\hat \sigma_e$ of planar two-phase strongly inhomogeneous systems
with different structures of inhomogeneities in magnetic field
\ci{12}. Two of them describe systems with various real "bulk"
inhomogeneities (i.e. having real 2D bulk inhomogeneities): the
compact inclusions  of regions of different sizes and forms
("random droplets") of one phase into another  and the "random
parquet" structure constructed from square plaquettes with randomly
distributed stripes of two phases \ci{13}.  The third one has the effective conductivity, obtained by a dual
 "magnetic" transformation from the known effective medium
approximation (EMA) formula, based on the wire-network
representation of inhomogeneous systems \ci{6} (we will name it the effective medium
(EM) model).

The corresponding effective conductivities have the following forms
$$
\sigma_{ed} (\{\sigma\}, \{x\}) =
\fr{\sigma'_{ed}(ac+b)}{(\sigma'_{ed})^2 + a^2}, \quad
\sigma_{et} (\{\sigma\}, \{x\}) = c \fr{(\sigma'_{ed})^2 -a
b'}{(\sigma'_{ed})^2 + a^2}, 
\en(10)
$$
where $\sigma'_{ed}$ is the effective conductivity of the models without magnetic field, but it depends on transformed partial arguments 
$\sigma'_{id} = \sigma_{id}/\sigma_{ai}$ with $\sigma_{ai} = \sigma_{it} + a.$
The parameters of the "magnetic" transformation  $a,b'=b/c,c$
depend on the partial conductivities and have the following form
$$
a_{\pm} = \fr{|\sigma_2|^2 - |\sigma_1|^2 \pm
\sqrt{B}}{2(\sigma_{1t} - \sigma_{2t})}, \quad b'_{\pm} =
\fr{|\sigma_1|^2 - |\sigma_2|^2 \pm \sqrt{B}}{2(\sigma_{1t} -
\sigma_{2t})},\quad c=-a,
$$
$$
B = [(\sigma_{1t} - \sigma_{2t})^2 + (\sigma_{1d} -
\sigma_{2d})^2] [(\sigma_{1t} - \sigma_{2t})^2 + (\sigma_{1d} +
\sigma_{2d})^2], 
\en(11)
$$
where $|\sigma_i|^2 = \sigma_{id}^2 + \sigma_{it}^2,$ and,
evidently, $B\ge 0.$
The effective conductivities $\sigma_{ed} (\{\sigma\}, \{x\})$ of the three
inhomogeneous models without magnetic field have simple forms:
\bs

(1) a "random
droplets" model
$$
\sigma_{ed} (\{\sigma\}, \{x\})= (\sigma_{1})^{x_1}(\sigma_{2})^{x_2};
\en(12)
$$

(2) a "random parquet" model

$$
\sigma_{ed} (\{\sigma\}, \{x\}) =
\left(\sigma_{1} \sigma_{2} \right)^{1/2} \left(\fr{x_1
\sigma_{1} + x_2 \sigma_{2}}{x_1
\sigma_{2} + x_2 \sigma_{1}}\right)^{1/2};
\en(13)
$$

(3) an effective medium  model

$$
\sigma_{ed} (\{\sigma\}, \{x\}) = (x-\fr{1}{2})
(\sigma_{1} - \sigma_{2}) + \sqrt{ (x-\fr{1}{2})^2 (\sigma_{1} - \sigma_{2})^2 +
\sigma_{1} \sigma_{2}}.
\en(14)
$$

All formulas (10,12-14) correctly reproduce boundary values of the
effective conductivities as well as their exact values at equal
phase concentrations $x=1/2$ \ci{12}
$$
\sigma_{ed}(x=1/2)=\sqrt{\sigma_{1d} \sigma_{2d}} A, \quad
A = \left[1+\left(\fr{\sigma_{1t} - \sigma_{2t}}{\sigma_{1d} + \sigma_{2d}}\right)^2 \right]^{1/2},
$$
$$
\sigma_{et}(x=1/2)=\fr{\sigma_{2t} \sigma_{1d} + \sigma_{1t} \sigma_{2d}}{ \sigma_{1d} + \sigma_{2d}}.
\en(15)
$$
This permits us to write the exact expression
for $R$ at $x=1/2$ 
$$
R(1/2,H)= 
\sigma_{ed}(1/2,0) \fr{\sigma_{ed}(1/2,H)} {\sigma_{ed}^2(1/2,H) + \sigma_{et}^2(1/2,H)} - 1,  
$$
where
$$
\sigma_{ed}(1/2,H)=\fr{\sqrt{\sigma_{10} \sigma_{20}}}{\sqrt{(1+H^2)(1+\eta^2 H^2)}} A(H),
$$
$$
A(H)= 
\left[1+H^2\left(\fr{(1+\eta^2 H^2)- \eta z(1+H^2)}{(1+\eta^2 H^2)+ z(1+H^2)}\right)^2 \right]^{1/2},
$$
$$
\sigma_{et}(1/2,H)=\fr{\sigma_{20}(1+\eta)H}{(1+\eta^2 H^2)+ z(1+H^2)}.
\en(16)
$$
One can obtain an asymptotic behaviour of $R$ at high $H, \eta H \gg 1.$ 
Since in this limit
$$
\sigma_{ed}(1/2,H)=\fr{\sqrt{\sigma_{10} \sigma_{20}}}{ H}
\left|\fr{\eta - z}{\eta^2 + z}\right|,
\quad
\sigma_{et}(1/2,H)=\fr{\sigma_{20}(1+\eta)}{(\eta^2 + z) H},
$$
then
$$
R(1/2,H)= c H + o(H), \quad c = \fr{|\eta -z|(\eta^2+z)}{(\eta -z)^2 + z(1+\eta)^2}  + o(H).
\en(16')
$$
Thus, the exact value $R(1/2,H)$ has a linear dependence on $H$ at high $H$ with the coefficient $c,$ which becomes 0  for homogeneous media as well as for inhomogeneous systems with $\eta = z.$ It is interesting that $c=1$  at $\eta=-1.$ Just in this case $\sigma_{et}(x=1/2)=0.$ At $\eta=1$ $c$ reduces to a simple formula
$c =\fr{1-z}{1+z}.$

Substituting (11) and (12-14) into (10) and then into (6), one can
obtain the explicit expression for the magnetoresistance $R(x,H)$
of the corresponding systems at arbitrary concentrations and magnetic fields. Since these formulae are
rather complicate we will analyze their behaviour by constructing
the plots of their dependencies on $x$ and $H$.

Using  explicit formulae (6),(7),(10-14) we have constructed the
plots of $H$- and $x$-dependencies of $R(x,H)$ for inhomogeneous
systems, whose inhomogeneity structures are similar to three
considered models, for different values of the inhomogeneity parameter $z$
and $\eta$. The corresponding plots of $H$- and $x$-dependencies at $z=10^{-2}$ and $\eta = 1$ are represented, respectively,
 in Fig.1 and Fig.2 ((a)-(c)).

\begin{figure}[t]
\begin{tabular}{cc}
\input epsf \epsfxsize=5.5cm \epsfbox{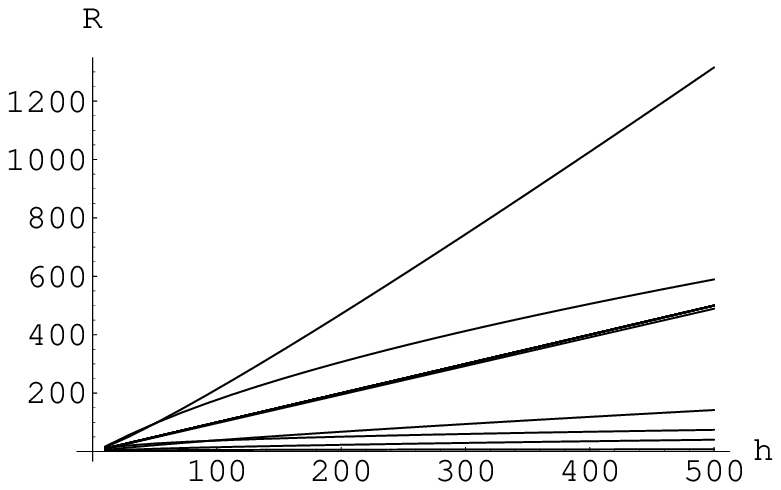}&
\input epsf \epsfxsize=5.5cm \epsfbox{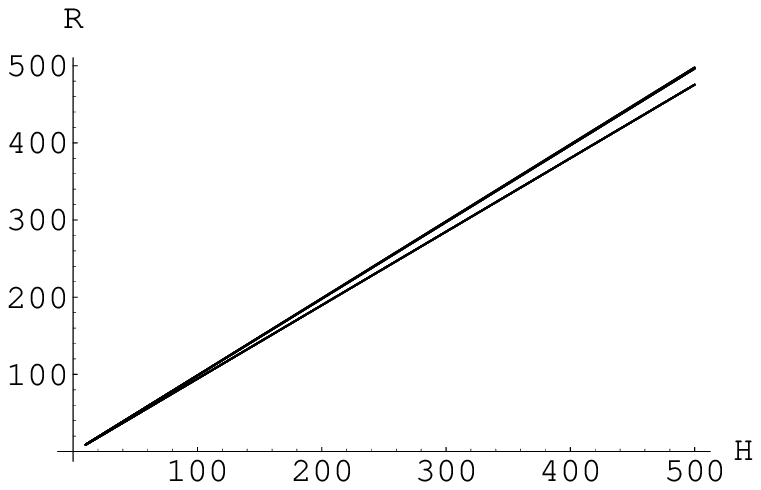}\\
{} & {}\\
(a) & (b)\\
\end{tabular}

\vs{0.3cm}

\begin{tabular}{c}
\input epsf \epsfxsize=5.5cm \epsfbox{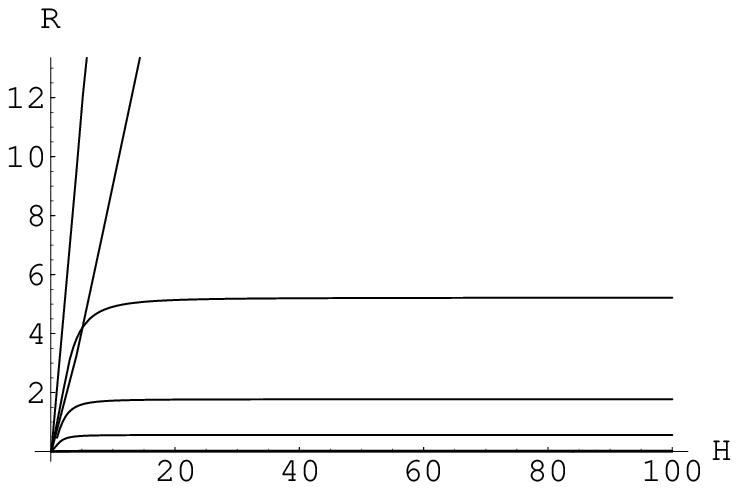}\\
{}\\
(c)\\
\end{tabular}

\vs{0.3cm}

{\small  Fig.1.  The  plots of the $H$-dependence of the
magnetoresistance $R(x,H)$ for three explicit expressions obtained
above (respectively, (a),(b),(c)) at the inhomogeneity parameters
$z=0.01,$ and $\eta = 1$ at some characteristic values
of concentration $x.$ 
}
\end{figure}
One can see from fig.1 and fig.2  that the behaviour of the
magnetoresistance, though different for various expressions (i.e. various models)
and saturated in some regions of $x$, has two common characteristic features:

1) absolute values of $R$ are very large, especially at peaks;

2) all values in the regions with large values of $R$
increase with a growth of $H$ at relatively high $H$
approximately linearly. 
 
Both these properties are the consequencies of the high $H$ behaviour of the effective conductivities \ci{12} 
$$
\sigma_{ed} \sim \sigma_{et} \sim 1/H,
\en(17)
$$
and are qualitatively consistent with the experimental results from
\ci{2,14,15}, which show a large magnetoresistivity and its
approximately linear growth with an increase of $H.$ 
Practically the same plots for $R(x,H)$ are obtained for the case, when two partial phases have carriers with opposite charges, i.e. at $\eta=-1.$ It follows also from
(6), since $\sigma_{et},$  sensitive to a sign of $\eta,$ enters in $R$ only as squared. Some dependence on a sign of $\eta$ appears for the Hall resistivity (see
next paper).

More detailed analysis shows a number of interesting specific properties. Let us start with
the "random droplets" model. In this case $R$ has evident one peak asymmetrical structure positioned for the inhomogeneity parameter $z=10^{-2}$ in the region of concentration $x \sim 0.37$ with
a maximum growing almost linearly and slightly shifting to the higher values of concentration and with a width narrowing with an increase  of $H.$ Note that the maximal value of $R$ is much larger then the exact value (15). It appears that
a decline of linear dependence also changes becoming maximal $\approx 2.7$ at the peak and $H=300.$ For higher values of inhomogeneity parameter $z$ the peak slightly
shifts to the lower values of $x$ (for $z =10^{-3}$ it is situated at $x \approx 0.3$).
\begin{figure}[t]
\begin{tabular}{cc}
\input epsf \epsfxsize=5.5cm \epsfbox{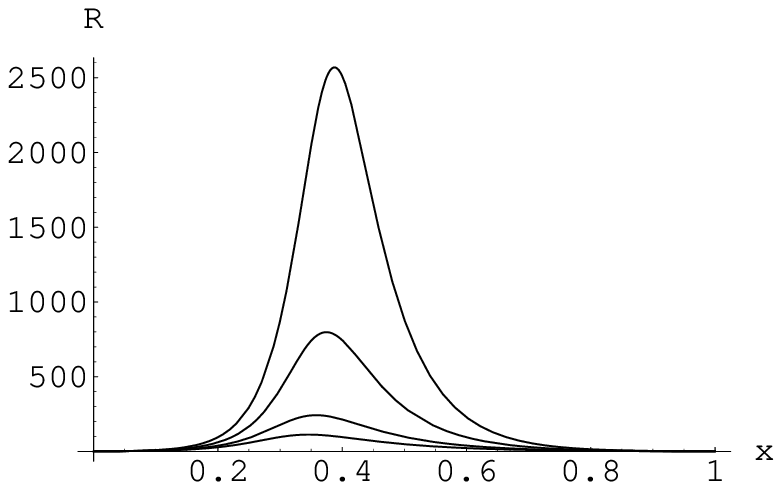}&
\input epsf \epsfxsize=5.5cm \epsfbox{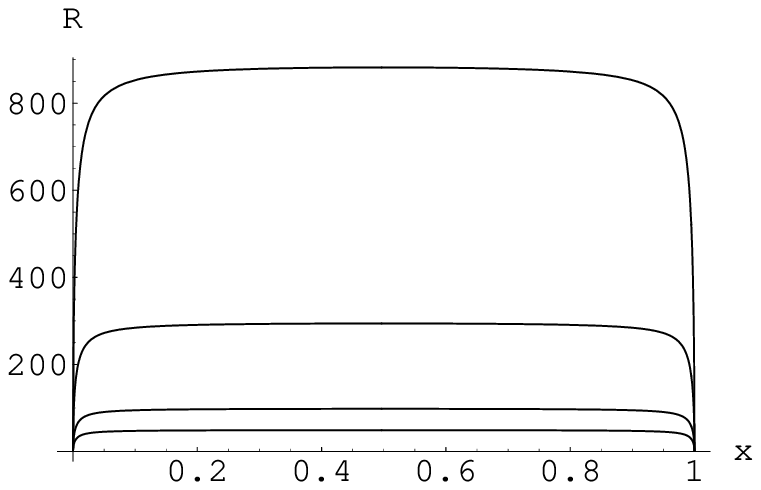}\\
{} & {}\\
(a) & (b)\\
\end{tabular}

\vs{0.3cm}

\begin{tabular}{cc}
\input epsf \epsfxsize=5.5cm \epsfbox{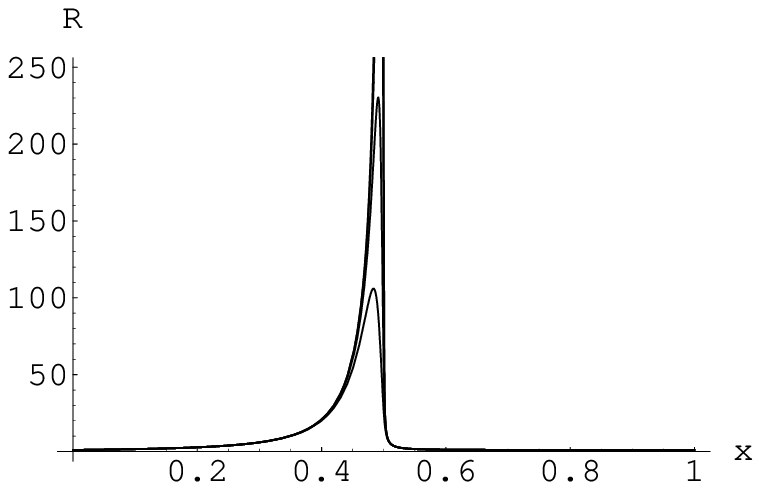}&
\input epsf \epsfxsize=5.5cm \epsfbox{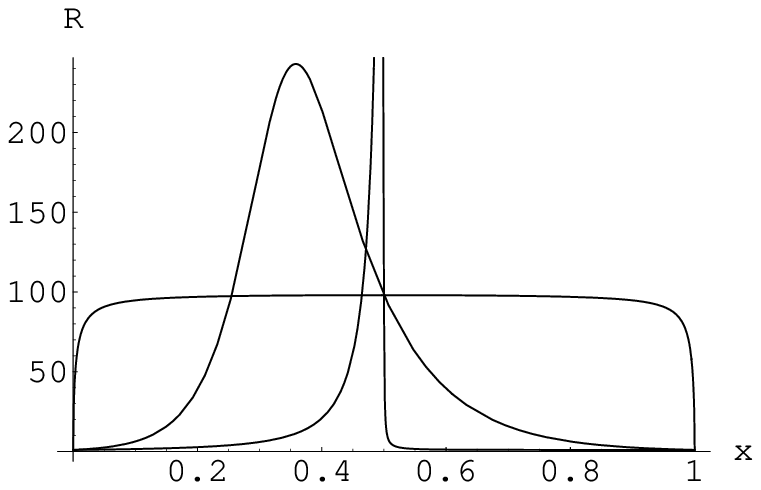}\\
{} & {}\\
(c) & (d)\\
\end{tabular}
\vs{0.3cm}

{\small  Fig.2.  The  plots of the $x$ dependence of the
magnetoresistance $R(x,H)$ for three explicit expressions obtained
above (respectively, (a),(b),(c)) at the inhomogeneity parameter
$z=0.01,$ and at the four different
(dimensionless) values of magnetic field $H:$ 1) 50, 2) 100,  3)
300 and 4) 900 (the corresponding plots go from the lower to the
upper ones; on (c) $H$ = 25, 50, 100, since in this case for higher $H$ the peaks are very narrow and almost coincide). (d) shows a comparison of all $R(x,H)$ at $H=100.$}
\end{figure}

Very remarkable pictures are obtained for the "random parquet" model. The magnetoresistance shows approximately constant large values, almost coinciding with the exact value at $x=1/2$ in a wide region of concentrations and very sharp changes in small regions $x, 1-x \ll 1$ near the boundary concentrations $x=0,1.$
In this model the exact value is a maximal one. The width of the plateau region, which is approximately symmetrical, slightly grows with a growth of $H.$ For the concentrations in the plateau region $R$ increase linearly with $H$ with an approximately constant coefficient, which coincides with its exact value at $x=1/2$ from (16). The width  of the narrow regions near boundaries decreases with an increase of $H$ and a decrease of $z.$

For the effective medium model we have obtained very unusual pictures.
$R(x,H)$ shows strongly asymmetrical very narrow and high peak, situated just below $x=1/2,$
with a sharp decrease to negligably small values in the narrow region near $x=1/2$
and long small tale in the region $x < 1/2.$ The exact value of $R$ is significantly smaller than the maximal one.

In order to compare $R(x,H)$ for different models in more detail we
represent also the plots of their $x$-dependencies at
$H = 50$ in one Fig.2(d).
One can see from Fig.2(d) the relative widths and maximal values of $R$ for three
models. The highest values $R$ achieves for the "random droplet" and effective medium models, meanwhile the "random parquet" model has the widest region of relatively large values. At equal phase concentrations $x=1/2$ all models give exact value (16,16'). 
\bs

\und{\bf 4. Discussion and comparison with the experiments}

\bs
The obtained formulae and the constructed plots unambiguously show an existence of the large 
magnetoresistance effect with almost linear growth at relatively large magnetic fields in various planar inhomogeneous models. In different models it has  different dependences on concentration. The analysis of these  dependences and the structures 
of the considered models allows to conclude that the different inhomogeneous fluctuations have different influence on the magnetoresistance. The random compact  (droplet like) fluctuations have the most strong influence in the range of concentrations $0.3 \le x \le 0.5.$  In this region the randomly distributed 
droplets of higher conducting component serve as traps for charge carriers. They work
the most effectively for concentrations in the region $0.3 \le 0.4$ where all droplets have finite sizes and rather rare intersections.  For larger concentrations the large droplets appear. Though  the carriers are 
confine  inside them for a long time, they already
can propagate on significant distancies due to their large size. The last effect induces a decrease of $R.$ The almost linear growth of $R$ with $H$ means that the
stronger magnetic fields help to confine carriers inside droplets.  

At the same time, the stripe-like inhomogeneities, as it follows from the results
for  the "random parquet" model, begin to work as traps already at very small concentrations $x$, when they have a form of rare, randomly distributed, stripes of highly conducting component. If a length of these stripes is constrained and they do not form any effectively large or dense cluster (what takes place in this model due to a construction), these
stripes gives approximately the same contribution to $R,$ very weakly depending on their concentration. Only in very narrow regions near $x=0$ and $x=1,$ where a
number of highly or badly conducting stripes is very small, the rare randomly distributed stripes give sharp changes of $R,$ which quickly  saturate.
We suppose that a similar behaviour will take place in the models with other randomly distributed stripe-like inhomogeneities. For example, we expect that such behaviour (maybe, a more smoother than in the original "random parquet" model) will
take place in models constructed from plaquettes of different forms and sizes with
isotropic discrete (or even continuous) distributions of stripe orientations. 

The most unusual behaviour shows the "magnetically" transformed effective medium
model. It does not influence on $R$ almost for all concentrations, except very 
narrow region near $x=1/2$ and long tail with non large values in the region $0 \le
x \le 0.4.$ It can be explained by the tight connection of this model with the wire-type networks, which cannot be used for a description of  magneto-transport effects, connected with bulk properties. Only in the narrow region near $x = 1/2,$
where the large dense clusters with many "dead"  ends appear, they give strong
contribution to $R.$ 

Now we can compare the considered models with the known results for silver chalcogenides ($Ag_{2+\delta}Se$ and $Ag_{2+\delta}Te$), which show large linear
magnetoresistance effect (see, for example, \ci{2,14}) for different concentrations 
of $Ag$ \ci{15}.
As it was noted in \ci{2} silver chalcogenides have a tendency to a formation of
different phases on small scales (including nanometre one). The additional $Ag$ ions 
strive to situate on various defects and grain boundaries \ci{15,16}. The latter presumably have a form of intersecting straight lines. When concentration of the excess $Ag$ ions is small, they form randomly distributed stripes of different size along these boundaries \ci{15}. This structure is very similar to the corresponding structure of the "random parquet" model. The dependences of $\rho_{e0}$ and $R(x,H)$ on $x$ in the "random parquet" model at small $x$ (the $x$-dependences of $\rho_{e0}$ in all three considered above models can be 
found, for example, in \ci{13}, where they are represented for corresponding $\sigma_{e0}$, one only needs to change $\sigma_1, \sigma_2$ into $\rho_2$ and $\rho_1$ respectively) are also very similar to the behaviour of $\rho_{e0}$ and $R(x,H)$ observed in experiments \ci{2,15}, where very small excess of $Ag$ ions gives sharp increase of $R.$
Moreover, for the larger concentrations, in some finite region, the experiment shows an approximately constant behaviour of $\rho_{e0}$ and $R,$ which changes again on, respectively, decreasing and increasing behaviour \ci{15}. We can explain this behaviour as a result of a crossover from the presumably striped structure of excess $Ag$ ions to the mixed structure, including also their compact islands \ci{15}. The latter forms a structure, which is similar to that of the "random droplets" model. The corresponding contribution into $R$ is firstly small
relative to "random parquet" model. But, as follows from our Fig.2(d) (its envelope in the region $0 \le x \le 0.5$ qualitatively coincides with Fig.3 from \ci{15}), at approximately $x_c=0.25$ (the exact values of the crossover $x_c$ can depend on magnetic field and inhomogeneity parameters) a contribution of the "random droplets" model becomes larger than that of the "random parquet" model and again appears the increasing behaviour of $R$ (or a decreasing behaviour of $\rho_{e0}$).
It means that the structure of inhomogeneities with the excess $Ag$ ions in $Ag_{2+\delta}Se(Te)$ is very similar to the structures of our models (at least, in the region $0 \le x \le 0.5$) and the latter can give the necessary formulae for a
description of the magneto-transport properties of silver chalcogenides in this region of concentrations. Moreover, our formulae give also numerical values, which
are in a good agreement with the experimental ones (see our forthcoming papers).

At the end we would like to note a possible importance of the obtained results
for practical applications. As is known, a behaviour of $R$ such as it takes place in the "random" parquet model and models related with it is the most suitable for
magnetic sensors and information storage technologies (see, for example, \ci{17}).
For this reason, the "random parquet" type models with random stripe-like structures of inhomogeneities show us what kind of inhomogeneous structure one needs to fabricate sensors and other hightechnology devices, having small sizes and working at room temperatures.
\bs

\und{\bf 5. Conclusion}
\bs

Using three explicit approximate expressions for the effective
conductivity of 2D isotropic two-phase systems in a magnetic
field, we have obtained explicit expressions for 
the magnetoresistance $R(x,H)$
of planar isotropic two-phase
systems. The  plots of the $x$- and $H$-dependencies of $R(x,H)$ 
at the different values of the inhomogeneity parameters
$z$ and $\eta$  are constructed. They show the evident large, almost linear,
magnetoresistance effect. The behaviour of the constructed plots is very similar and qualitatively compatible with the experimental data for silver chalcogenides $Ag_{2+\delta}Se$ from \ci{2,15}. 
A possible physical explanation of such behaviour is proposed. It is noted that the
random, stripe type structures of inhomogeneities, are the most suitable for a fabrication
of magnetic sensors and a storage of information at room temperatures.
We hope that the constructed models and their modifications as well as the obtained
results  can be applied also for a description of magneto-transport properties of various real heterophase systems (regular and nonregular as well as random),
satisfying the symmetries and having the similar structures, in a wide
range of inhomogeneity parameters and at arbitrary concentrations
and magnetic fields. 

\bs
\und{ Acknowledgments}

\bs
The authors are thankful to A.A.Abrikosov, P.Littlewood and M.Parish for very useful
information about their works.
This paper is 
supported by the RFBR grants 02-02-16403, and by ESF network AQDJJ, EPSRC grant
GR/S05052/01, the
Royal Society  grant 2004/R4-EF and Ministry of Science (UK).

\end{document}